\documentstyle[preprint,aps]{revtex}
\draft
\begin{document}               
\title{
Ground states of a system of interacting particles in a parabolic trap
}
\author{M. S. Hussein and  O.K. Vorov
}
\address{
Instituto de Fisica, 
Universidade de Sao Paulo \\
Caixa Postal 66318,  05315-970,  \\
Sao Paulo, SP, Brasil
}
\date{13 June 2001}
\maketitle
\begin{abstract}
Within the ``lowest Landau level approximation'',
we develop a method to find the ground state of a 2d system
of interacting particles confined by a parabolic potential.
\end{abstract}

\newpage

The problem of finding exact eigenstates of a 2d system of 
particles interacting by the two-body central forces $V(r)$ and
confined by an external parabolic potential arises in many subfields,
including cooled and trapped fermionic atoms, Bose-Einstein condensates 
in traps, quantum dots and electrons in magnetic field.
Assuming spherical symmetry,
we are interested here in the lowest energy states of such system
as a function of exact quantum numbers.
The lowest Landau level approximation is equivalent to 
the weak interaction limit 
\begin{displaymath}
V(r) \ll \hbar \omega
\end{displaymath}
where $\omega$ is the
oscillator frequency of the external trapping potential.
The effective Hamiltonian 
of the $N$-body system
is the sum of the energy of degenerate level\cite{1}
with total angular momentum $L$ 
($\hbar$$=$$m$$=$$1$) 
\begin{equation}\label{ham}
H= \omega (L+N)+{\tilde V}=
\quad 
P(L) \left[
\sum_i^{N}
\left(\vec{p}_i^2+\omega^2\vec{r}_i^2\right)/2 +
\sum\limits_{i>j} 
V(r_{ij}) \right] P(L)
\end{equation}
which comes from the noninteracting Hamiltonian 
in the 2d parabolic trap,
and the projected interaction ${\tilde V}$.
Here, $P(L)$ is the projector onto the space of lowest Landau level, $S$.
The space $S$ is formed by the homogeneous polynomials in  
$z$$_i$$=$$x$$_i$$+$$i$$y$$_i$
of degree $L$ times the Gaussian factor 
$|0$$\rangle$$=$$e$$^{-\frac{1}{2}\sum|z_k|^2}$.
Hereafter, we set $\omega$$=$$1$ for simplicity.
The polynomials are antisymmetric (symmetric) for identical
fermions (bosons). The eigenstates of the $H$ (\ref{ham})
are found from diagonalization of $\tilde{V}$ in the space $S$.

The effective
interaction ${\tilde V}$ can be found in the form \cite{2}
\begin{equation}  \label{OPE}
{\tilde V}= 
\sum_{k=0}^{[L/2]} 
V_k,
\qquad
V_k\equiv s_{2k}(B^{2k}-B^{2k-1})+
(s_{2k+2}-s_{2k+1})B^{2k+1},
\end{equation}
where 
$B^k=\sum\limits_{j>i}^{N}(a^{\dagger}_i-a^{\dagger}_j)^k
(a_i-a_j)^k$ and
$a^{\dagger}_i=z_i/2-\partial/\partial z^*_i$ and 
$a_i=z^*_i/2+\partial/\partial z_i$ are the standard ladder operators.
Here, 
$B^0$$\equiv$$N$$(N$$-$$1$$)$$/$$2$ and $B^{-1}$$\equiv$$0$;
$[k]$ denotes the integer part of $k$.  
In (\ref{OPE}), 
$s$$_k$$=$$\int_0^{\infty}$$\frac{dt}{k!}$$M$$($$k$$+$$1$$,$$1$$,$$-$$t$$)$
$V$$($$\sqrt{2t}$$)$
with $M$ the Kummer function.

In order to find the ground state without solving the 
whole spectrum, we use the following idea\cite{2}.
Suppose that the Hamiltonian 
can be written as a sum 
\begin{equation}\label{SUSY}
\tilde{V}= V_0 + V_S, \qquad (i) \quad V_S|0)=0, \qquad (ii)\quad V_0|0) = E_{min}|0), 
\end{equation}
such that 
the second term (i) annihilates a state $|0)$, 
and (ii) $|0)$ is the ground state for the first term
(this ground state may be degenerate).
If (iii) the second term
is non-negative definite (i.e., it has no negative eigenvalues)
\begin{equation}\label{POSITIVITY}
\quad V_S \geq 0,
\end{equation}
then
the state $|0)$
will still be the ground 
state 
of the full Hamiltonian $V_0+V_S$, with 
the same eigenvalue ${E}_{min}$\cite{2}.

In the case of Bose system, 
the polynomials of the basis states are
$S_L=P_S m_L P_S$, where 
$m_L=z_1^{l_1}z_2^{l_2}...z_N^{l_N}$ is the monomial
with 
a given partition of the integer 
$L=\sum\limits_n l_n$, and 
$P_S$ is 
the operator of symmetrization over $N$ variables.
The above decomposition (\ref{SUSY},\ref{POSITIVITY}) 
with the properties (i) and (ii)
can be found by inspecting action of (\ref{OPE}) on the basis states. 
The simplest basis state 
\begin{equation}\label{simplest}
|0_L) = P_S z_1 z_2 ... z_L |0\rangle , \qquad (\tilde{V}-V_0)|0_L)=0,
\end{equation}
is annihilated by any term in (\ref{OPE}),
except for $V_0$.
The term $V_0$ can be written as
\begin{equation}\label{V0}
V_0 = ( N / 2 ) [ ( N - 1 ) s_0 - ( L - v )
( s_1 - s _2 ) ] .
\end{equation}
Here, 
\begin{displaymath}
v=\frac{1}{N} \sum\limits_{j,k}^{N} a^+_j a_k
\end{displaymath} 
is the collective contribution to the
angular momentum, which is also a conserved quantum number, $[v,\tilde{V}]=0$.
The allowed values of $v$ are $0,1,2,..,L-2,L$\cite{2}.
The state (\ref{simplest}) can be represented as a sum
\begin{equation}\label{EXPANSION}
|0_L) = \sum_{v=0}^{N} |0_{L,v}), 
\qquad
|0_{L,v})= Z^v P_S \prod_{k=1}^{L-v}(z_k-Z) |0\rangle,
\end{equation}
where $Z$$=$$\sum_{k=1}^{N}$$z$$_k$$/$$N$ is the collective variable.
Each term $|0_{L,v})$ is the eigenvector of $v$ and therefore of $V_0$.
Each $v$-sector is represented
by a single state $|0_{L,v})$ in expansion (\ref{EXPANSION}).
The spectrum of $V_0$
consists of $L$ 
equidistant 
degenerate levels.
Conditions (i) and (ii) are therefore satisfied for the states 
$|0_{L,v})$.
In particular, this means that those are eigenstates of
$\tilde{V}$ with eigenvalues given by (\ref{V0})\cite{2},
see also \cite{3}.

If the term $V_S\equiv\tilde{V}-V_0$ is non-negative definite,
the condition (iii) is satisfied and each state 
$|0_{L,v})$ will be the {\it ground state} in corresponding sector $L,v$.
In fact, it is possible to find a broad universality class of 
the potentials $V(r)$ which 
give $V_S\equiv\tilde{V}-V_0\geq0$.
We work in the space $S$ and use 
permutation symmetry to
write $P_S V_S P_S$ in the form
\begin{equation}\label{INERTIA}
P_S \sum\limits_{i>j}V_{S,ij}  P_S 
= 
\frac{N(N-1)}{2}
P_S 
V_{S,12}P_S = \frac{N(N-1)}{2} P_S p_{12}V_{S,12} p_{12} P_S 
\end{equation}
where $V_{S,ij}$ is the contribution from pair of particles $i$$,j$
to $V_S$ (\ref{OPE},\ref{SUSY}), 
and $p_{ij}$ is the operator of symmetrization over this pair.
The condition $p_{12}V_{S,12} p_{12} \geq 0$ is a {\it sufficient condition}
for $P_S V_S P_S \geq 0$\cite{2}. The eigenvalues of 
$p_{12}V_{S,12} p_{12}$ can be easily evaluated, using the full 
space of monomials $m_L$.
As a result, the condition $p_{12}V_{S,12} p_{12} \geq 0$ 
reduces to the 
following 
set of integral inequalities
\begin{equation}\label{RESULT1}
\int\limits_0^{\infty}dt V(r) e^{-r^2/2}
\left[ \frac{r^{4n}}{2^{2n}(2 n)!} - 1 + n\left(1 -
\frac{r^4}{8}\right) \right] \geq 0  ,
\end{equation}
which must be satisfied for integer $n$, $0\leq 2 n \leq L$.
These conditions describe the broad universality 
class of interaction potentials
$V(r)$, for which the lowest energy states in sectors with
quantum numbers $v$ and $L$$\leq$$N$ are given by 
$|0_{L,v})$ [see Eq.(\ref{EXPANSION})], 
and their energies are equal to the
eigenvalues of $V_0$, Eq.(\ref{V0}), 
with $s$$_0$$=$$\int_{0}^{\infty}$$dt$$e^{-t}V(\sqrt{2t})$,
$s$$_1$$-$$s$$_2$$=$$\int_{0}^{\infty}$$dt e^{-t}$
$\left(\frac{1}{2}-\frac{t^2}{4}\right)$$V(\sqrt{2t})$.
Conditions (\ref{RESULT1}) are valid, in particular, for 
the Gaussian interaction $V$$($$r$$)$$=$
$U$$_0$$\frac{e^{-r^2/R^2}}{\pi R^2}$,$U_0$$\geq$$0$,
for the $\delta$-function interaction $V$$=$$U$$_0$
$\delta$$($$\vec{r}$$)$,$U_0$$\geq$$0$  and for the $2d$ log-Coulomb
interaction $V$$=$$U$$_0$$log$$($$1$$/$$r$$)$, $U$$_0$$\geq$$0$
and for many others potentials, which are predominantly repulsive.
These results are obtained for the Bose case.
Generalization of the method for the Fermi case is possible.

The work was supported by FAPESP.

\end{document}